\documentclass[boxit]{JAC2003}

\usepackage{graphicx}
\usepackage{booktabs}
\usepackage{verbatim}
\usepackage{subcaption}
\begin{document}

\title{Current Status of the SANAEM RFQ Accelerator Beamline}

\author{G. Turemen\thanks{gorkem.turemen@ankara.edu.tr}, B. Yasatekin,  Ankara University, Ankara, Turkey.\\ S. Ogur, V. Yildiz, Bogazici University, Istanbul, Turkey.\\ O. Mete, Cockcroft Institute and University of Manchester, UK.\\ S. Oz, A. Ozbey, H. Yildiz, Istanbul University, Istanbul, Turkey.\\ F. Yaman, IZTECH, Izmir, Turkey. \\ Y. Akgun, A. Alacakir, S. Bolukdemir, TAEK, SANAEM, Ankara, Turkey.\\ A. Bozbey, A. Sahin, TOBB ETU, Ankara, Turkey.\\ G. Unel, University of California at Irvine, Irvine, California, USA. \\S. Erhan, University of California at Los Angeles, LA, California, USA.}

\maketitle

\begin{abstract}

The design and production studies of the proton beamline of SPP, which aims to acquire know-how on proton accelerator technology thru development of man power and serves as particle accelerator technologies test bench, continue at TAEK-SANAEM as a multi-phase project. For the first phase, 20 keV protons will be accelerated to 1.3 MeV by a single piece RFQ. Currently, the beam current and stability tests are ongoing for the Inductively Coupled Plasma ion source. The measured magnetic field maps of the Low Energy Beam Transport solenoids are being used  for matching various beam configurations of the ion source to the RFQ by computer simulations. The installation of the low energy diagnostics box was completed in Q1 of 2015. The production of the RFQ cavity was started with aluminum 7075-T6 which will be subsequently coated by Copper to reduce the RF (Ohmic) losses.  On the RF side, the development of the hybrid power supply based on solid state and tetrode amplifiers continues. All RF transmission components have already produced with the exception of the circulator and the power coupling antenna which are in the manufacturing and design phases, respectively. The acceptance tests of the produced RF components are ongoing. 
This work summarizes the design, production and test phases of the above-mentioned SPP proton beamline components.

\end{abstract}

\section{Introduction}

\begin{figure}
   \centering
   \includegraphics*[width=75mm]{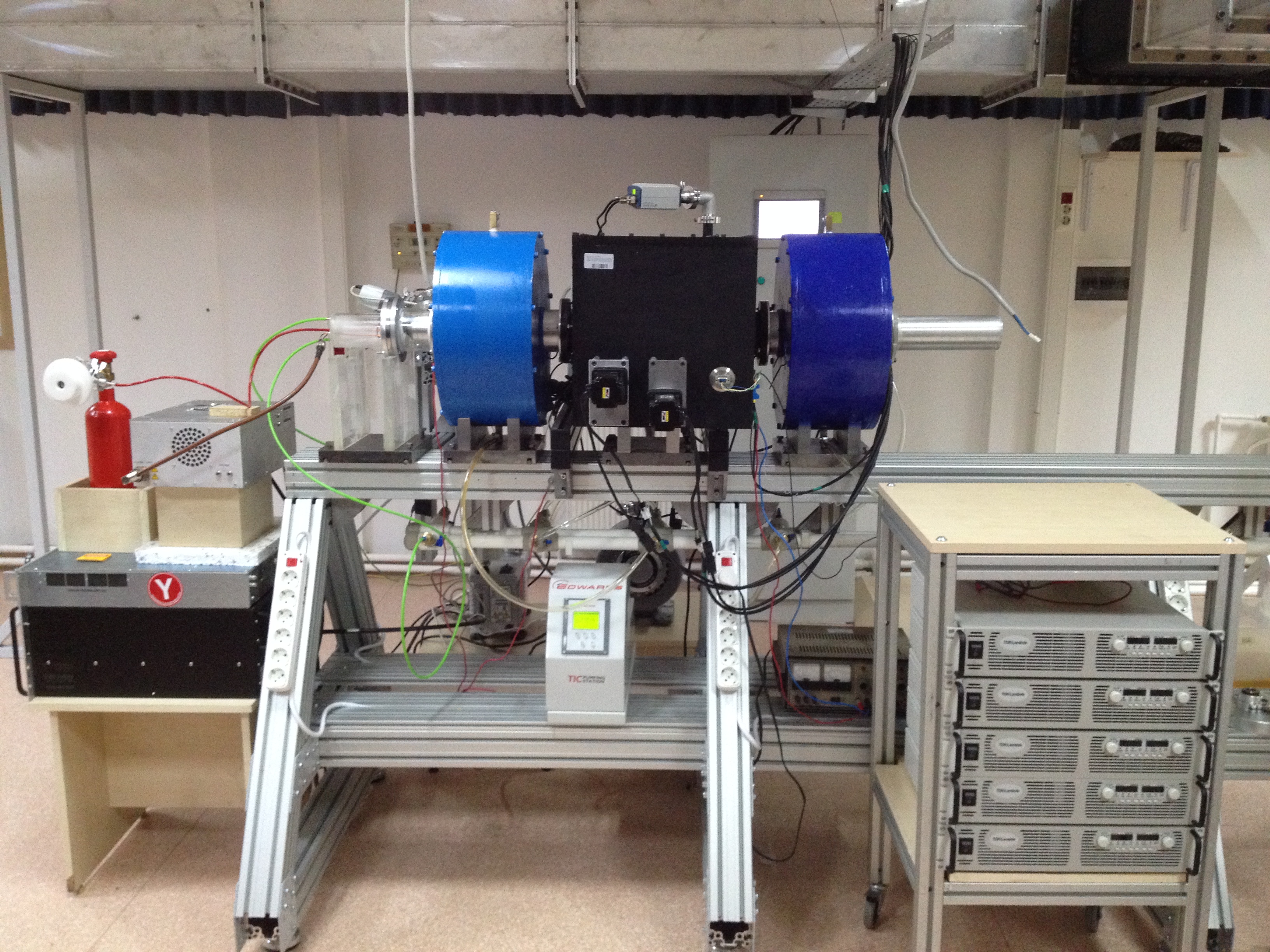}
   \caption{Currently installed parts of the SPP beamline.}
   \label{beamline}
\end{figure}

Turkish Atomic Energy Authority (TAEK) started the construction of a proton beamline at its Saraykoy Nuclear Research and Training Center, SANAEM. The SANAEM Prometheus Project (SPP) is designed and engineered with local resources enabling accelerator physicists to acquire know-how and it will serve as a particle accelerator technologies test bench. Currently some of the components are being manufactured (e.g. the single piece RFQ) or being assembled (e.g. the hybrid RF PSU). In the meantime, the stability tests and measurements continue with the already installed components such as the ion source, the LEBT line and the low energy diagnostics box (Fig.~\ref{beamline}). The rest of this note details the ongoing work on the components of the SPP beamline.

\section{Ion Source}

An Inductively Coupled Plasma (ICP) type ion source was selected  for its low production cost and its ability to match the relatively modest proton current requirements of the SPP project. Initially, a water cooled plasma chamber with 8 multicusp magnets was produced with an internal RF antenna \cite{ipac14}. Later, plasma generation method was changed because of the high voltage sparking between the alumina coated copper internal RF antenna and the high voltage plate. In the new design, a quartz tube with a gas injection nose is used as the plasma chamber. A 3.5 turn external RF antenna is used to produce hydrogen plasma and to prevent sparking. This RF antenna is sandwiched between the quartz plasma chamber and the plexi-glass water cooling chamber. With this setup, the ICP ion source is found to be suitable for longtime operation without overheating at an RF power below 500 W. The impedance matching between the RF antenna and the 13.56 MHz RF power supply was achieved using a matching network. The RF power coupling to the plasma was easily accomplished for a chamber diameter of 30 mm at 10$^{-3}$ mBar. 

\subsection{Extraction Region and Beam Current Tests}

The extraction system of the ion source consists of a plasma electrode and a puller electrode with 3 mm and 5 mm apertures, respectively. The plasma electrode should be kept at 20 keV due to the specific proton beam energy requirement for the proper RFQ operation. The matching between the plasma meniscus and electric potential can be optimized by adjusting the plasma to puller electrode distance. The emittance evolution for different plasma to puller electrode distances ($Dp$) for several current densities ($J$) is studied with IBSimu \cite{ibsimu} simulations. The rms emittance of the proton beam is significantly reduced from 0.14 to 0.02 $\pi$.mm.mrad for a distance of 15 mm at all investigated current densities. At the beam current tests, by setting the plasma to puller electrode distance to 15 mm and for 400 W RF power, the measured ion current just after the source is 3.5 mA. More precise characterization of the proton beam such as profile, emittance and current measurements will be performed at the low energy beam diagnostics box.

\section{LEBT Line}

SPP beamline's LEBT components are already produced and installed, except two axis steerer magnets which are in the manufacturing phase. In addition, a home-made ACCT will be mounted after the Pulse Forming Network's installation on the plasma electrode to generate a pulsed beam.   

\subsection{Magnetic Field Maps of Solenoids}

\begin{figure}
   \centering
   \includegraphics*[width=74mm]{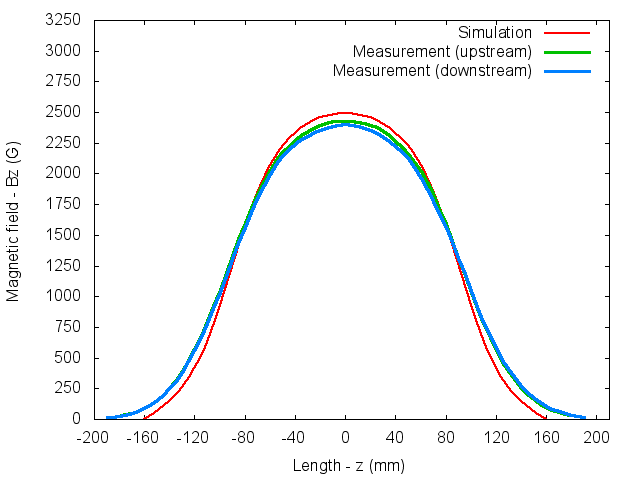}
   \caption{ LEBT solenoids magnetic field maps.}
   \label{sol}
\end{figure}

The LEBT line has two water-cooled solenoid magnets. These identical magnets are designed to generate 0.0447 T.m integrated field on the beam axis at 14 A coil current. The magnetic field maps of the solenoids are measured with 5 mm steps by using axial hall probe at the beam axis. After the coil current stability reached by water cooling, the integrated fields are measured as 0.0459 T.m and 0.0453 T.m for upstream and downstream solenoids, respectively. The integrated field errors compared (Fig.~\ref{sol}) to the POISSON \cite{poisson} simulations are found as \%2.59 and \%1.29. Due to the relatively small deviations from the measurements and the lack of a 3D field map, the simulated field map was used for the LEBT optimizations. The optimized solenoid current values  will be scaled for the real operation according to measured field deviations from simulations. 

\subsection{RFQ Acceptance Match}

\begin{figure}
   \centering
   \includegraphics*[width=70mm]{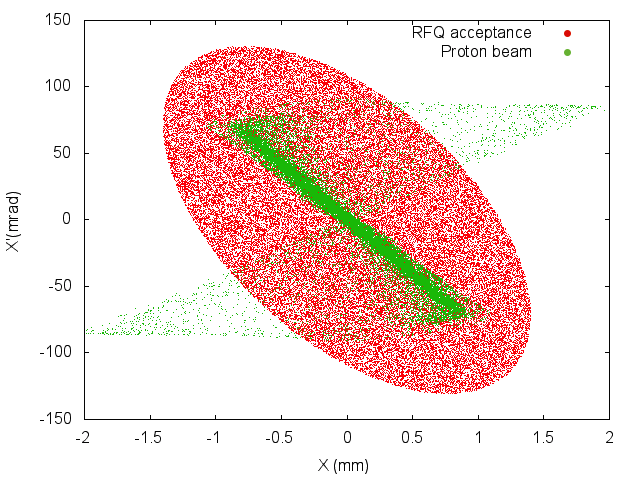}
   \caption{The acceptance ellipse of the designed RFQ (red) and matched proton beam  (green).}
   \label{match}
\end{figure}

TRAVEL \cite{travel} is used to estimate the coil currents of the solenoids and the drift space lengths of LEBT to match the beam phase space to the RFQ acceptance. By using DELTA \cite{delta} for the consecutive TRAVEL simulations, the optimized coil currents and the drift lengths are found for the different ion source configurations. As can be seen in Fig.~\ref{match},  the proton beam is matched to the RFQ acceptance for the selected (J=500 A/m$^2$ and Dp=15 mm) ion source beam with 80 cm length between the solenoids and 15 cm distance between the downstream solenoid and the RFQ plate.

\subsection{Low Energy Beam Diagnostics Box}

A low energy beam diagnostics box is designed and produced to measure the proton beam characteristics such as current, profile and emittance. The beam current is measured by a simple Faraday Cup (FC) with a suppressor electrode. The FC is designed by using FLUKA \cite{fluka} and SIMION \cite{simion} codes. The former is used to estimate back-scattering proton quantity from the cup material which is aluminum and to optimize the cup shape and length. The latter is used to calculate the required value of the negative voltage to prevent secondary electrons from escaping the cup. 19 cm length and $-150$ V  potential are enough for recollect the back-scattering protons and the electrons, respectively. After design studies the FC was produced and installed in the diagnostics box. A phosphor screen is used to measure the transverse beam shape and to calculate transverse emittance by incorporating a pepper-pot foil. As seen in Fig.~\ref{beamline}, the diagnostics box is installed  and is ready for the beam tests.

\section{RFQ}

The SPP RFQ is a four-vane type 352.2 MHz cavity. As reported before \cite{linac14} the beam dynamics and the 2D electromagnetic (EM) designs were completed using LIDOS.RFQ \cite{lidos} and SUPERFISH \cite{poisson}, respectively. After 2D simulations, CST MWS \cite{cst} is used for the 3D EM design. Although the SPP RFQ's resonant frequency is 352.2 MHz, the RFQ was designed for 350 MHz in 2D and 351.2 MHz in 3D to ensure cavity tunability. All  mechanical requirements such as cooling channel spaces and machining capabilities of the vane production were considered during design.

\subsection{MWS Simulations}

The 3D EM simulations and the optimization studies are performed with CST MWS's eigenmode solver. The CAD models of the modulated vane tips based on points exported from LIDOS.RFQ are imported to MWS to perform simulations. Other cavity regions such as vane skirts, cutbacks, tuners, vacuum and RF ports are modeled directly on MWS.

\subsection{Cutback Design}

\begin{figure}
   \centering
   \includegraphics*[width=70mm]{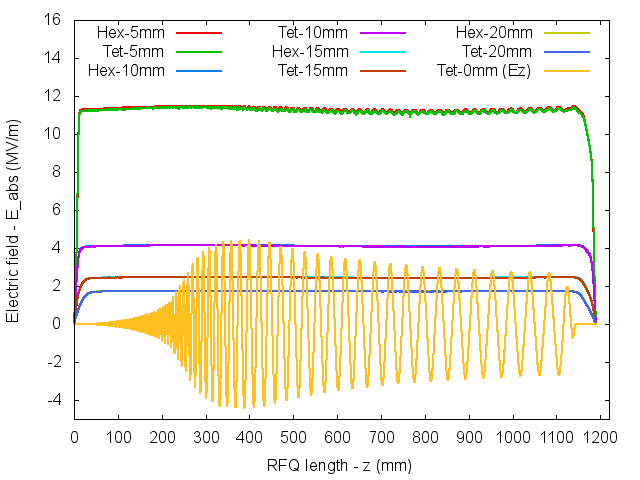}
   \caption{The scaled electric field on the different distances away from the beam axis in both transverse directions of the untuned RFQ.}
   \label{efield}
\end{figure}

Parametric MWS simulations are performed to provide field flatness along the cavity,  considering the operating frequency and the quality factor. Detailed optimization simulations are performed with JDM solver by using three different vacuum inserts. These vacuum inserts are meshed with different mesh sizes which are increasing from the beam axis to cavity wall. After all optimizations, the untuned field flatness of \%2.5 is found acceptable (Fig.~\ref{efield}) for the RFQ which oscillates in a proper resonant frequency by taking into account meshing artifacts. The tetrahedral mesh solver was used by importing the final RFQ CAD model to  MWS to cross-check the production geometry (Fig.~\ref{efield}). After cavity production, the field flatness will be fine-tuned by using 16 identical tuners to ensure proper operation of the RFQ.

\subsection{Tuner Effects}

\begin{figure}
   \centering
   \includegraphics*[width=70mm]{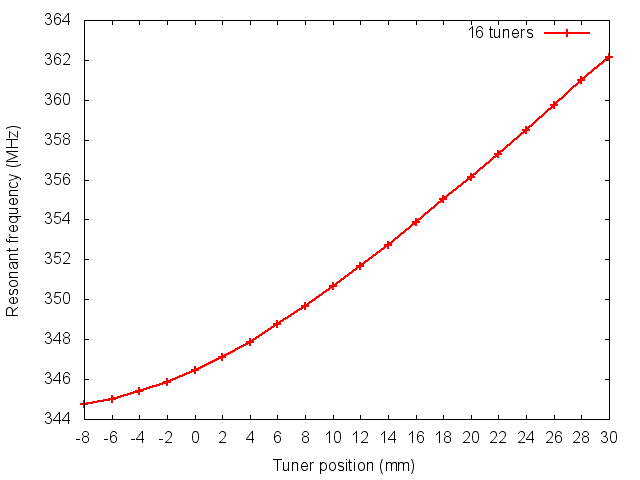}
   \caption{The resonant frequency shift of the RFQ with the different tuner positions.}
   \label{tuner}
\end{figure}

For the 1.3 m cavity, 16 of 57 mm diameter tuners are designed except RF and vacuum ports which  can also be used. The tuners have +30 mm and -8 mm tuning range from the flushed position to the cavity wall. The frequency tuning range of the tuners is estimated by using a RFQ model without modulation. After parametric MWS simulations, the frequency tuning ranges of the tuners are found to be adequate as shown in Fig.~\ref{tuner}.

\subsection{Vane Manufacturing}

\begin{figure}
   \centering
   \includegraphics[width=73mm]{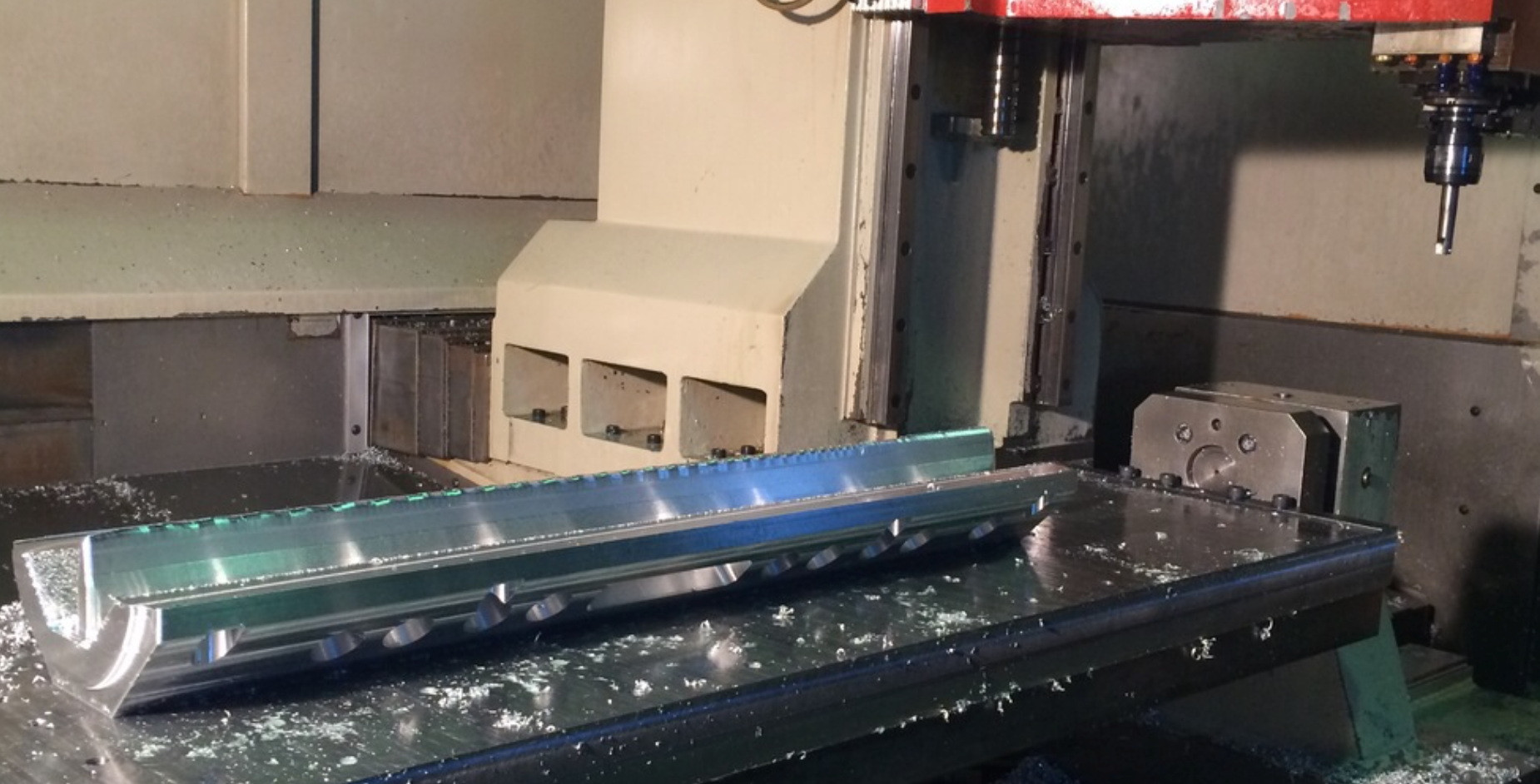}
   \caption{Test machining of the RFQ vane.}
   \label{testrfq}
\end{figure}

The SPP RFQ engineering design followed beam dynamics and EM simulations. As a result of four-vane design, a couple of two identical vanes should be produced. Before final production, a machining test was done with a reduced (\%30) dimension vane. A photo of the produced vane can be seen in Fig.~\ref{testrfq}. The final vane machining started just after the CMM results of the test vane were found to be acceptable.

\section{Outlook}

The  SPP beamline work is progressing smoothly, without any problems foreseen in the future. The local manufacturing abilities have proven to be adequate for all components produced so far. The production of the most critical component, the RFQ cavity has just started. If there are no delays in production and commissioning, the first accelerated protons are expected by early 2016.

\section{Acknowledgments}
The authors are grateful to A. Tanrikut for useful comments. This study is funded by TAEK with a project No. A4.H4.P1.

\end{document}